\documentclass{aa}  

\usepackage[colorlinks=true,allcolors=blue]{hyperref}

\usepackage{threeparttable}
\usepackage[rightcaption]{sidecap}

\usepackage{graphicx}
\usepackage{txfonts}
\usepackage{xspace}
\newcommand{\kms}{km~s$^{-1}$\xspace}
\newcommand{\ergs}{erg~s$^{-1}$\xspace}
\newcommand{\Msunyr}{M$_{\odot}$~yr$^{-1}$\xspace}
\newcommand{\Msun}{M$_{\odot}$\xspace}

\newcommand{\hb}{H$\beta$\xspace}
\newcommand{\ha}{H$\alpha$\xspace}
\newcommand{\heii}{He\,{\sc{ii}}\xspace}

\newcommand{\naid}{Na\,{\sc{id}}\xspace}

\newcommand{\oiii}{[O\,{\sc{iii}}]\xspace}

\newcommand{\sii}{[S\,{\sc{ii}}]\xspace}

\newcommand{\lya}{Ly$\alpha$\xspace}
\newcommand{\civ}{C\,{\sc{iv}}\xspace}
\newcommand{\ciii}{C\,{\sc{iii}}]\xspace}

\newcommand{\nii}{[N\,{\sc{ii}]}\xspace}

\newcommand{\GS}{GS10578\xspace}

\begin{document}

   \title{GA-NIFS: AGN activity in a Ly$\alpha$ emitter within a triple-AGN system anchored by a passive galaxy at $z=3$}

   \author{Michele Perna
          \inst{\ref{iCAB}}\thanks{e-mail: mperna@cab.inta-csic.es}
          \and
          Santiago Arribas\inst{\ref{iCAB}}
          \and
          Mahmoud Hamed\inst{\ref{iCAB}}
          \and
          Francesco D'Eugenio\inst{\ref{iKav},\ref{iCav}}
          \and
          Andrew~J.~Bunker\inst{\ref{iOxf}}
          \and
          Stefano~Carniani\inst{\ref{iNorm}}
          \and
          St\'ephane~Charlot\inst{\ref{iSor}}
          \and
          Roberto~Maiolino\inst{\ref{iKav},\ref{iCav},\ref{iUCL}}
          \and 
          Bruno Rodr\'iguez~Del~Pino\inst{\ref{iCAB}}
          \and
          Hannah~\"{U}bler\inst{\ref{iMPE}}
          \and
          Torsten~Böker\inst{\ref{iESAusa}}
          \and
          Elena~Bertola\inst{\ref{iOAA}}
          \and
          Giovanni~Cresci\inst{\ref{iOAA}}
          \and
          Isabella~Lamperti\inst{\ref{iUNIFI}, \ref{iOAA}}
          \and
          Giacomo~Venturi\inst{\ref{iNorm}}
          \and
          Michele~Ginolfi\inst{\ref{iUNIFI}, \ref{iOAA}}
          \and 
          Montserrat~Villar~Martín\inst{\ref{iCAB}}
          \and
          Sandra~Zamora\inst{\ref{iNorm}}
          }

   \authorrunning{M. Perna et al.}
   \titlerunning{GA-NIFS: a $z = 3$ low-mass \lya emitter hosting an AGN}

   \institute{
            Centro de Astrobiolog\'ia (CAB), CSIC--INTA, Cra. de Ajalvir Km.~4, 28850 -- Torrej\'on de Ardoz, Madrid, Spain\label{iCAB}
    \and
            Kavli Institute for Cosmology, University of Cambridge, Madingley Road, Cambridge, CB3 0HA, UK\label{iKav}
    \and
            Cavendish Laboratory - Astrophysics Group, University of Cambridge, 19 JJ Thomson Avenue, Cambridge, CB3 0HE, UK\label{iCav}
    \and
            Department of Physics, University of Oxford, Denys Wilkinson Building, Keble Road, Oxford OX1 3RH, UK\label{iOxf}
    \and
            Scuola Normale Superiore, Piazza dei Cavalieri 7, I-56126 Pisa, Italy\label{iNorm}
    \and
            Sorbonne Universit\'e, CNRS, UMR 7095, Institut d’Astrophysique de Paris, 98 bis bd Arago, 75014 Paris, France\label{iSor} 
    \and
            Department of Physics and Astronomy, University College London, Gower Street, London WC1E 6BT, UK\label{iUCL}  
    \and
            Max-Planck-Institut f\"ur extraterrestrische Physik (MPE), Gie{\ss}enbachstra{\ss}e 1, 85748 Garching, Germany\label{iMPE}
    \and    
            European Space Agency, c/o STScI, 3700 San Martin Drive, Baltimore, MD 21218, USA\label{iESAusa}
    \and 
            INAF - Osservatorio Astrofisico di Arcetri, Largo E. Fermi 5, I-50125 Firenze, Italy\label{iOAA}
    \and
            Università di Firenze, Dipartimento di Fisica e Astronomia, via G. Sansone 1, 50019 Sesto F.no, Firenze, Italy\label{iUNIFI}
             }

   \date{Received September 15, 1996; accepted March 16, 1997}

 
  \abstract
   {Massive quenched galaxies at $z>3$ challenge models of early galaxy evolution, as their rapid formation and abrupt quenching require efficient feedback, often linked to active galactic nuclei (AGN). The quiescent galaxy \GS at $z=3.064$ is a key example of this population. Previous JWST/NIRSpec IFU observations revealed an AGN-driven outflow and uncovered a compact pair of AGN separated by $\sim5$~kpc. In addition, VLT/MUSE spectroscopy has identified a third AGN candidate at a projected distance of $\sim30$~kpc, associated with a luminous \lya emitter (LAE2) characterised by high-ionisation UV lines, although rest-frame optical diagnostics were not previously available.
   }
   {We aim to confirm the nature of LAE2 using rest-frame optical diagnostics enabled by new JWST observations, and to characterise the physical and ionisation properties of both LAE2 and a distinct nearby \lya emitter (LAE1) that lacks any detectable continuum counterpart. Through this analysis, we investigate the interplay between low-mass satellites, black-hole growth, and the ionised environment of a quenched massive galaxy at high redshift.}
   {We analyse new NIRSpec IFU observations targeting the optical nebular lines of LAE1 and LAE2, including \hb, \oiii, \ha, and \nii, complemented with VLT/MUSE spectroscopy, as part of the GA-NIFS project. We extract integrated and spatially resolved spectra, construct emission-line maps, and use standard diagnostic diagrams to determine ionisation sources and kinematics.
   }
   {LAE2 exhibits rest-frame UV–optical line ratios fully consistent with an embedded AGN. Its \oiii and \ha emission displays a clumpy morphology and irregular, non-rotating kinematics on sub-kpc scales. Except for \lya, LAE1 remains undetected in all nebular lines and in JWST imaging; its \lya emission is broad ($\sigma \sim 200$~\kms) and asymmetric. The similarity of the LAE1 and LAE2 \lya profiles in both velocity and flux suggests that LAE1 traces resonantly scattered emission powered by LAE2 rather than in-situ star formation (although the latter cannot be completely ruled out).  
   }
  {Our analysis reveals that the environment of \GS contains both multi-black-hole activity and gas structures on tens-of-kpc scales, offering new insights into how feedback, accretion, and satellite interactions influence the late evolutionary stages of quenched massive galaxies.}


   \maketitle
%

\section{Introduction}

Rapid galaxy formation and subsequent cessation of star formation (quenching) in massive galaxies shortly after the Big Bang ($z\gtrsim3$, 1--2 Gyr before cosmic noon) pose a major challenge to models of galaxy evolution (e.g. \citealt{Glazebrook2017, Glazebrook2024, Boylan-Kolchin2023,  Carnall2023MNRAS, Valentino2023, Baker2025, deGraaff2025NatAs, McConachie2025, Weibel2025}). Understanding the mechanism responsible for removing cool gas (i.e. the fuel for star formation) or preventing its supply, is crucial. Current theory favours cumulative feedback from central supermassive black holes (SMBHs), which can act either through ejective mechanisms (removing gas via outflows; e.g. \citealt{KingPounds2015, Xie2024sim}) or preventative mechanisms (stopping fresh accretion; e.g. \citealt{Croton2006, Zinger2020sim, Bluck2022}). 

The James Webb Space Telescope (JWST), particularly with the Near-InfraRed Spectrograph (NIRSpec) in its integral field spectroscopic (IFS) mode (\citealt{Boker2022,Jakobsen2022}), has begun to probe these monumental galaxies during this key evolutionary epoch (e.g. \citealt{PerezGonzalez2025NatAs, Pascalau2025}).
The galaxy \GS at $z = 3.064$ is an excellent laboratory for studying quenching processes. It is a highly massive system (M$_\star \approx~10^{11}$~\Msun) classified as a quiescent galaxy, with a current star-formation rate (SFR) estimated to be $\le6$~\Msunyr (\citealt{DEugenio2024NatAs,Scholtz2024gs10578}). \GS hosts an active galactic nucleus (AGN), confirmed by its X-ray detection ($L_{\rm 2-10~keV} = 6.4\times 10^{44}$~\ergs) and extreme position on the ``Baldwin, Phillips \& Terlevich'' (BPT; \citealt{Baldwin1981}) emission-line ratio diagnostic diagram (\citealt{Perna2025dual}). 
It is classified as a radio-quiet AGN, with $q24 = 0.49 \pm 0.03$ (with $q24$ defined as  observed ratio of 24~$\mu$m to radio luminosity) detected both at 3 and 6 GHz ($4\times 10^{40}$ and $1\times 10^{40}$~\ergs, respectively; \citealt{Lyu2022demo}).

Direct evidence of AGN feedback in \GS has been established, demonstrating powerful neutral-gas outflows traced by \naid absorption, with mass outflow rates ($\approx 30-100$~\Msunyr) significantly exceeding the estimated SFR, strongly supporting the dominance of ejective feedback in quenching the galaxy (\citealt{DEugenio2024NatAs}; see also \citealt{Venturi2025gsagn}). Furthermore, \GS exhibits ordered stellar rotation, suggesting that quenching occurred without destroying the galaxy’s disk structure (\citealt{DEugenio2024NatAs}).
Non-detection of CO(3-2) in deep (7-hours integration) Atacama Large Millimeter/submillimeter Array (ALMA) observations surprisingly reveals that the galaxy has a cold molecular gas mass M$_{\rm mol}<~10^{9.1}$~M$_\odot$; this represents one of the most stringent upper limits on the gas mass for a quiescent galaxy at high redshift (\citealt{Scholtz2024gs10578}).

\begin{figure}[!t]
   \centering
   \includegraphics[width=0.48\textwidth]{{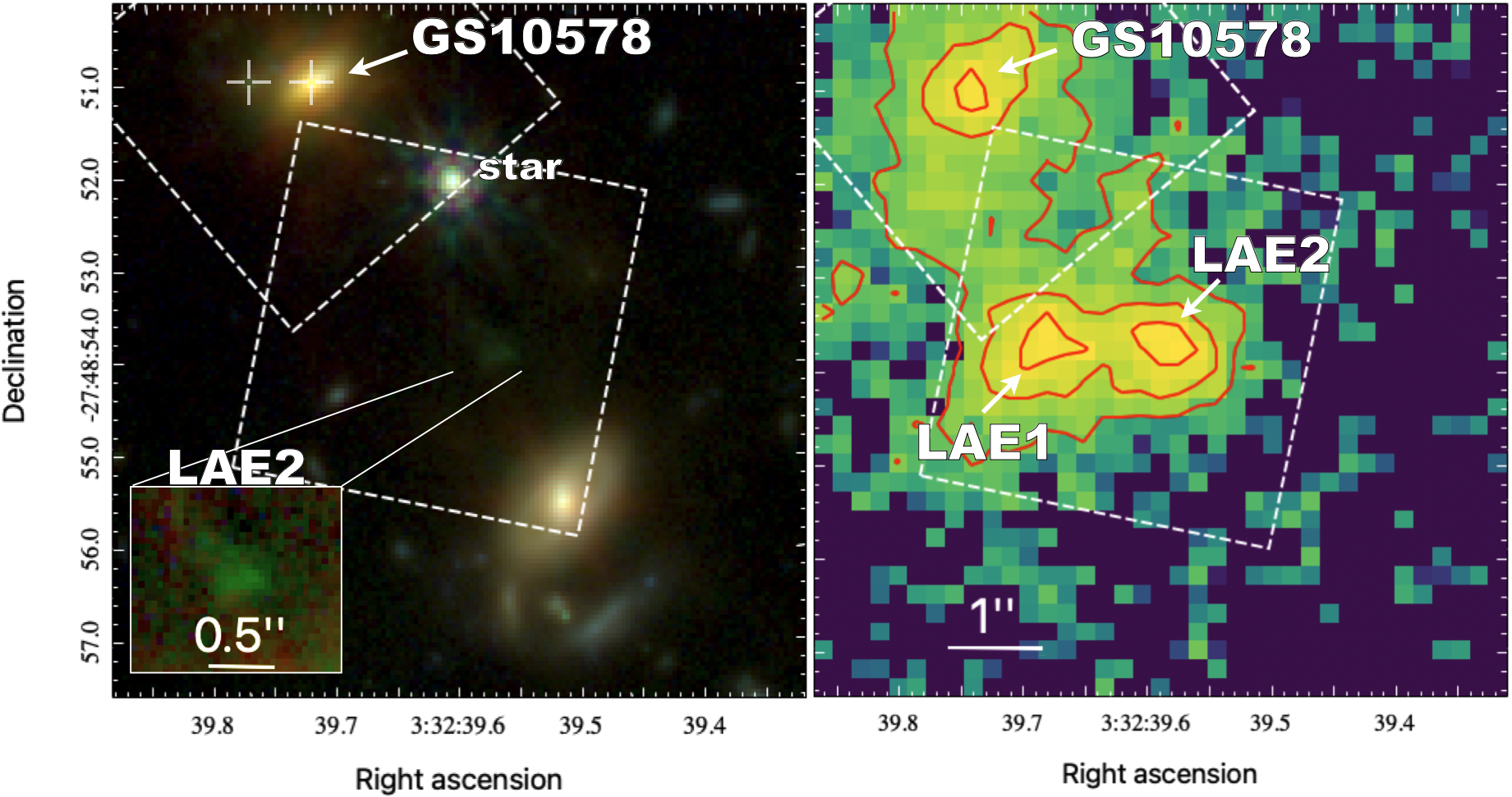}}
   \caption{
   Environmental context of \GS and its close emitters. {\it Left:} NIRCam three-colour composite image constructed from the F090W (blue), F200W (green), and F444W (red) filters. The red continuum emission from the main \GS galaxy is visible in the upper-left region, while the faint, clumpy \oiii emitter associated with LAE2 is detected near the field centre (see also the zoom-in inset). The two white crosses identify AGN-A (in \GS) and AGN-B (in the close satellite). Right: VLT/MUSE narrow-band image obtained by integrating over the \lya line (in the range [--1000,+600]~\kms with respect to the systemic velocity of \GS), revealing the extended nebula encompassing \GS, LAE1, and LAE2 (contours at 3, 6, and 9$\sigma$). The footprints of the NIRSpec IFS observations from Cycle 1 (covering \GS) and Cycle 3 (targeting LAE1 and LAE2) are overlaid in both panels. 1\arcsec\ corresponds to 7.8~kpc at $z \sim 3$.
   }
   \label{fig:3colours}%
\end{figure}

Recent JWST/NIRSpec IFS and JWST/NIRCam imaging, together with archival optical observations with IFS Multi Unit Spectroscopic Explorer (MUSE) at the Very Large Telescope (VLT), have revealed that \GS resides in a complex, dense environment, potentially harbouring a triple AGN system (\citealt{Perna2025dual}). In addition to the primary AGN in \GS (AGN-A) and a close companion that also hosts an active nucleus (AGN-B, 4.7 kpc separation) on the basis of optical line diagnostics, the system includes multiple gas-rich satellites detected in NIRCam medium-band filters covering the \oiii emission (\citealt{DEugenio2024NatAs}), and two \lya Emitters (LAEs) detected in VLT/MUSE data (\citealt{Perna2025dual}; see also Fig.~\ref{fig:3colours}). These two sources are situated at a projected separation of $\approx 30$~kpc from the nucleus of \GS, with a velocity offset of $\approx 200-300$~\kms from \GS. Notably, each of them has a \lya luminosity comparable to that of \GS ($\approx 10^{42}$~\ergs). 
We also note that \GS lies $\sim~50$~kpc (and $\sim~6000$~\kms) from another \lya halo discovered by \citet[][see e.g. Fig. 9 in \citealt{Perna2025dual}]{Leclercq2017}. This suggests that \GS is embedded in a large-scale overdensity.

Previous analysis of MUSE rest-frame UV spectra suggested that the source labelled LAE2, despite its low stellar mass (log(M$_\star$/M$_\odot$) $\approx 8.3 \pm 0.4$), is powered by an AGN based on UV line ratio diagnostics (e.g. \civ/\heii\!$\lambda$1640; \citealt{Perna2025dual}). LAE1 instead is only detected in \lya, not in any other UV lines nor in imaging data from HST, JWST/NIRCam, and JWST/MIRI observations (\citealt{Eisenstein2023}; \citealt{Rieke2023, Alberts2024SMILES}). Confirming the AGN nature of LAE2 is particularly relevant in the broader context of black-hole growth in low-mass galaxies and the assembly of multi-AGN systems at high redshift, which present key constraints for hierarchical galaxy formation models and early SMBH seeding scenarios.
The nature of LAE1, instead, provides complementary insight into the ionisation and gas structure of the environment, as its \lya-only detection may arise from resonant scattering or fluorescence powered by nearby sources within the system. Taken together, LAE1 and LAE2 can thus trace distinct physical processes in the surroundings of \GS, offering a unique opportunity to investigate how feedback, accretion, and satellite interactions shape the final stages of massive-galaxy assembly.

In this work, we present new, deep JWST/NIRSpec IFS observations specifically covering the rest-frame optical emission lines of LAE1 and LAE2: namely, \hb, \oiii\!$\lambda\lambda$4960,5008, \ha, \nii\!$\lambda\lambda$6550,6585, and \sii\!$\lambda\lambda$6718,6732. These two LAEs lie outside the NIRSpec IFS field of view (FOV) of the Cycle 1 observations used to characterise \GS (Fig.~\ref{fig:3colours}), and their optical line emission had not previously been accessible. Our primary objective is to robustly confirm the nature of LAE2 using established optical diagnostic diagrams, and to investigate its physical and kinematic properties in detail. We will also use available multi-wavelength information to investigate the nature of LAE1.
This paper is organised as follows: Section \ref{sec:observations} describes the observations and data reduction procedures;
Sect. \ref{sec:analysis} presents the data analysis; Sect.~\ref{sec:results} shows the ISM kinematic and physical properties inferred from emission line
analysis of NIRSpec IFS data. 
Finally, Sect.~\ref{sec:discussion} discusses the implications of our findings, and Sect.~\ref{sec:conclusions} summarises our conclusions.
Throughout, we adopt a \cite{Chabrier2003} initial mass function (IMF, $0.1-100$~M$_\odot$) and a flat $\Lambda$CDM cosmology with $H_0=70$~km~s$^{-1}$~Mpc$^{-1}$, $\Omega_\Lambda=0.7$, and $\Omega_m=0.3$.

\section{Observations and data reduction}\label{sec:observations}

The NIRSpec IFS data analysed in this work cover LAE1 and LAE2 and were observed with the JWST/NIRSpec spectrograph in its IFS mode \citep{Boker2022}. Data were obtained as part of the NIRSpec IFS GTO programme `Galaxy Assembly with NIRSpec IFS' (GA-NIFS\footnote{\url{https://ga-nifs.github.io/}}, e.g.~\citealt{Bertola2025,Jones2024b14, Lamperti2024,Trefoloni2025, Zamora2025egs}) under the programme ID 4528 (P.I. Kate Isaak).

The IFS observations were taken with the grating/filter pair G235H/F170LP. This results in a data cube with spectral resolution $R\sim2700$ over the wavelength range 1.7--3.1 $\mu$m \citep{Jakobsen2022}.
The observations were taken with the NRSIRS2 readout pattern (\citealt{Rauscher2017}) with 16 groups per integration and one integration per exposure, using a 12-point medium cycling dither pattern, resulting in a total exposure time of 3.9 hours, matching the one of Cycle 1 covering \GS. 

We downloaded raw data files from the Barbara A. Mikulski Archive for Space Telescopes (MAST) and subsequently processed them with the JWST Science Calibration pipeline (version
1.17.1) under the recommended Calibration Reference Data System (CRDS) context jwst\_1322.pmap\footnote{\url{https://jwst-crds.stsci.edu/display_build_contexts/}}. 
Some modifications to the pipeline allow us to improve the data quality and they are described in detail in \citet{Perna2023,Perna2025ring}. Here we summarise the main changes. First, we applied the ``calwebb\_detector1'' step of the pipeline to account for detector level correction. Before calibrating the count-rate images through the ``calwebb\_spec2'' module of the pipeline, we corrected them by subtracting the 1/f noise through a polynomial fitting.  We identified and removed outliers directly in the calibrated 2-d images, by applying an algorithm similar to ``lacosmic'' \citep{vanDokkum2001} as implemented by \cite{DEugenio2024NatAs}. The final data cube was created by combining the individual calibrated 2-D exposures by using the ‘drizzle’ weighting obtaining a cube with a spaxel size of 0.05\arcsec. 

The reduced MUSE data cube used in this work was retrieved from the ESO archive. The observations are part of the MUSE Ultra Deep Field (UDF) programme and correspond to the publicly released mosaic described in \citet{Bacon2017}. The full dataset consists of a $3\times3$ mosaic, with individual pointings reaching a total exposure time of $\sim10$ hours each. The publicly available data products have a spatial sampling of $0.2\arcsec$ per pixel and a FWHM$_{\rm PSF}\approx 0.7\arcsec$. The astrometric alignment of the JWST and MUSE data was computed relative to five bright stars detected in NIRCam images that have been registered to Gaia DR2.

The overall environmental context of the \GS system is illustrated in Fig.~\ref{fig:3colours}. We used JWST/NIRCam imaging obtained in the wide filters F090W, F200W, and F444W (as retrieved from the Cosmic Dawn JWST Archive, from the JADES programme PID1180), to construct the three-colour composite image in the left panel. The red continuum emission from \GS is clearly visible in the upper-left region of the field, while the central part of the image reveals faint, multi-component \oiii line emission associated with LAE2 (see zoom-in inset); LAE1 remains undetected in NIRCam images. The right panel of Fig.~\ref{fig:3colours} shows the corresponding VLT/MUSE map of the extended \lya emission, which spatially resolves LAE1 and LAE2 relative to the main system.

Both panels in Fig.~\ref{fig:3colours} indicate the FOV of the NIRSpec IFS observations with white dashed boxes. The Cycle~1 data cover the \GS galaxy, whereas the new Cycle~3 observations presented in this work target LAE1 and LAE2.

\section{Data analysis}\label{sec:analysis}

In this section, we describe the method used to derive the emission line properties of LAE1 and LAE2 by fitting the R2700 data. The method applies to both individual spaxel spectra and spatially integrated spectra. 

As the continuum is not detected in the spectra, we do not model it.
We fitted the most prominent gas emission lines by using the Levenberg-Marquardt least-squares fitting code CAP-MPFIT (\citealt{Cappellari2017}). 
In particular, we modelled the \ha, \hb, and \oiii\!$\lambda\lambda$4960,5008 lines. The \heii\!$\lambda4687$ line,  \nii\!$\lambda\lambda$6550,85 and \sii\!$\lambda\lambda$6718,32 doublets are not detected in our data. We apply a simultaneous fitting procedure, so that all line features of a given kinematic component have the same velocity centroid and intrinsic velocity dispersion (e.g. \citealt{Perna2020}). Intrinsic velocity dispersions are obtained using the JWST/NIRSpec fiducial resolving
power curves recorded here\footnote{\url{https://jwst-docs.stsci.edu/
jwst-near-infrared-spectrograph/nirspec-instrumentation/
nirspec-dispersers-and-filters.}}. 
Moreover, the relative flux of the \oiii doublet components was fixed to 2.99 (\citealt{Osterbrock2006}). Each fit was performed with one and two Gaussian components. The final number of kinematic components used to model individual spectra was derived on the basis of the Bayesian information criterion (\citealt{Schwarz1978}). For both integrated and spaxel-by-spaxel analysis a single Gaussian component is enough to reproduce the data.

\begin{table}[h]
\centering
\caption{Spatially integrated properties from the spectral analysis.}%
\begin{tabular}{|lc|}
\hline

\hline
LAE2: & \\
RA: 3:32:39.5;  DEC:  --27:48:53.9 & \\
\hline
\underline{Nuclear spectrum ($r = 0.15''$):} & \\
z & $3.06674_{-0.00005}^{+0.00004}$ \\ 
$\sigma_{\rm v,\ H\alpha}$ & $112_{-2}^{+5}$~\kms\\
log L(\oiii)/[\ergs\!] & $41.44\pm 0.04$ \\
log L(\ha)/[\ergs\!] & $41.24\pm 0.02$ \\
log L(\heii 1640)/[\ergs\!] & $40.82\pm 0.04$ \\

\ha/\hb & $2.82_{-0.25}^{+0.31}$\\
E(B--V) & $0.01_{-0.01}^{+0.10}$\\
log(\oiii/\hb) & $0.65_{-0.04}^{+0.04}$ \\
log(\nii/\ha) & $<-0.79$\\
log(\heii 4687/\hb) & $< -0.51$\\
log(\heii 4687/\hb)$^\dagger$ & $> -0.80$\\

\underline{Integrated spectrum ($r = 0.50''$):} & \\
log L(\lya)/[\ergs\!] & $41.81\pm0.02$\\
log L(\ha)/[\ergs\!] & $41.72\pm 0.10$ \\
$\Delta v_{\rm Ly\alpha}$ & $105_{-3}^{+5}$~\kms \\
$\sigma_{\rm v,~Ly\alpha}$& $176\pm 3$~\kms \\
$f_{\rm esc}$(\lya) & $0.15- 0.27$\\
EW(\lya) & 56$~\AA$\\

\hline\hline
LAE1: & \\
RA: 3:32:39.6;  DEC:  --27:48:53.8 & \\
\hline
\underline{Integrated spectrum ($r = 0.50''$):} & \\
log L(\lya)/[\ergs\!] &  $41.90\pm0.02$\\
log L(\ha)/[\ergs\!] & $<41.17$\\
log L(\ciii)/[\ergs\!] & $<40.53$\\
$\Delta v_{\rm Ly\alpha}$ & $184_{-4}^{+8}$~\kms\\
$\sigma_{\rm v,~Ly\alpha}$& $220_{-5}^{+7}$~\kms\\
$f_{\rm esc}$(\lya) & $\gtrsim0.64$\\
EW(\lya) & $\gtrsim140~\AA$\\

\hline

\end{tabular} 
\tablefoot{\lya-emitters properties from JWST/NIRSpec (rest-frame optical) and VLT/MUSE (rest-frame UV). Colour excess measured from \ha/\hb, assuming a \citet{Cardelli1989} dust attenuation law. \ha and \ciii upper limits for LAE1 obtained assuming a velocity dispersion of 112 and 210~\kms, respectively, corresponding to the values measured for the same lines in LAE2. LAE1 $\Delta v_{\rm Ly\alpha}$ relative to the LAE2 systemic zero-velocity. Luminosities are not corrected for extinction nor for aperture sizes. Equivalent widths are in rest-frame. All upper limits are quoted at the $3\sigma$ level.\\
$^\dagger$: assuming \heii\!$\lambda$1640/$\lambda$4687 = 6.47, see Sect.~\ref{sec:results}. }   
\label{tab:integratedproperties}
\end{table}

   \begin{figure*}[!h]
   \centering
   \includegraphics[width=0.95\textwidth, trim=0mm 0mm 0mm 0mm]{{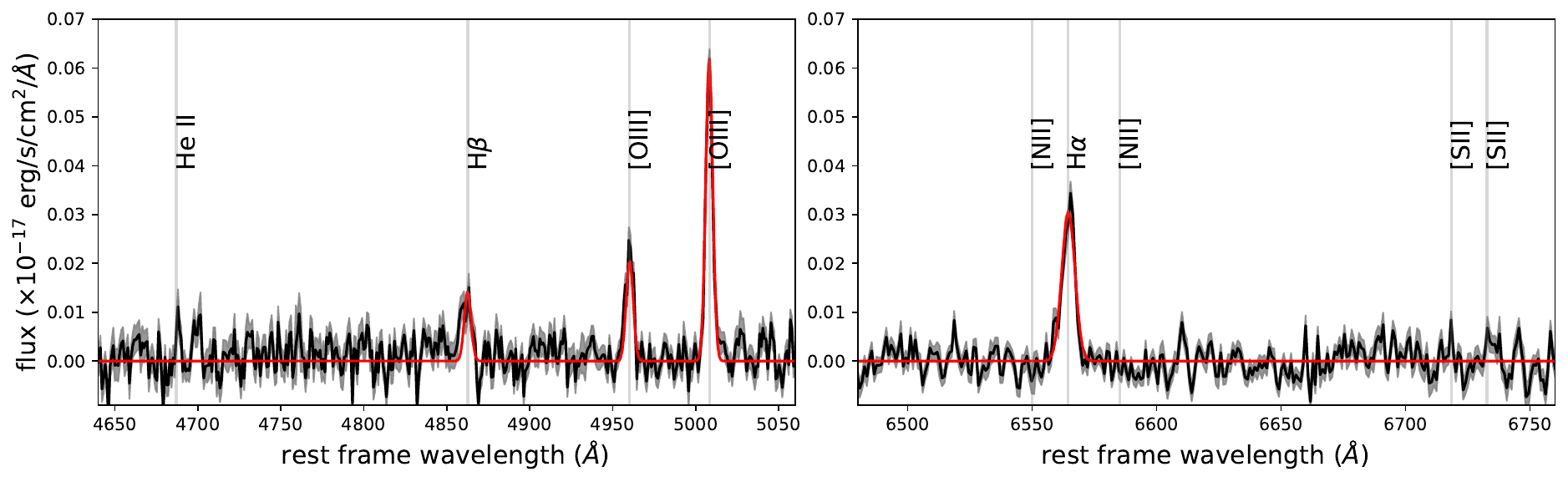}}

    \caption{Integrated NIRSpec spectrum of LAE2 extracted within a circular aperture of $r = 0.15''$. The main emission features are labeled, showing strong detections of \hb, \oiii\!$\lambda\lambda$4960,5008, and \ha. The \heii\!$\lambda$4687, \nii, and \sii lines are not detected above the $3\sigma$ noise level of the data.
    }
    \label{fig:integratedspectrum}%
    \end{figure*}

\section{Results}\label{sec:results}

The LAE2 shows multiple line detections, while no rest-frame optical emission is detected from LAE1 in the NIRSpec IFS data. This non-detection holds for all tested extraction apertures, from compact to extended, and is consistent with its lack of a counterpart in the NIRCam imaging.

Integrated spectra of LAE2 were extracted using circular apertures of radii $r = 0.15\arcsec$ and $r = 0.5\arcsec$. The smaller aperture maximises the signal-to-noise ratio (S/N) in the emission lines and is used to study the ionisation conditions in the nuclear region of LAE2. 
The current data, however, do not allow us to precisely identify the nucleus of LAE2. The highest-ionisation line detected, \heii\!$\lambda1640$, is observed with MUSE at low spatial resolution, while the NIRCam continuum imaging is not sufficiently deep for a robust morphological classification (Fig.~\ref{fig:3colours}). We therefore adopt the peak of the \oiii\!$\lambda5008$ emission as a proxy for the nuclear position of this system.
The larger aperture ($r = 0.5''$) captures the total emission from the system and enables a direct comparison with the \lya flux measured from MUSE data, for which the spatial resolution is insufficient to resolve sub-kpc scales within LAE2.

Figure~\ref{fig:integratedspectrum} presents the nuclear spectrum of LAE2. Prominent \oiii, \ha, and \hb emission lines are detected, while \heii, \nii, and \sii remain undetected. No continuum emission is observed. No clear evidence for a broad-line region is found, although the modest S/N of the \ha line (S/N~$\lesssim 15$) does not allow us to exclude its possible presence. The best-fit single-Gaussian model is shown in red in Fig.~\ref{fig:integratedspectrum}, and yields a systemic redshift of $z = 3.06674 \pm 0.00005$. The integrated fluxes and kinematic properties derived from both apertures are summarised in Table~\ref{tab:integratedproperties}. We note that the Balmer ratio is compatible with no extinction.

\begin{figure*}[!h]
   \centering
   \includegraphics[width=0.95\textwidth, trim=0mm 0mm 0mm 0mm]{{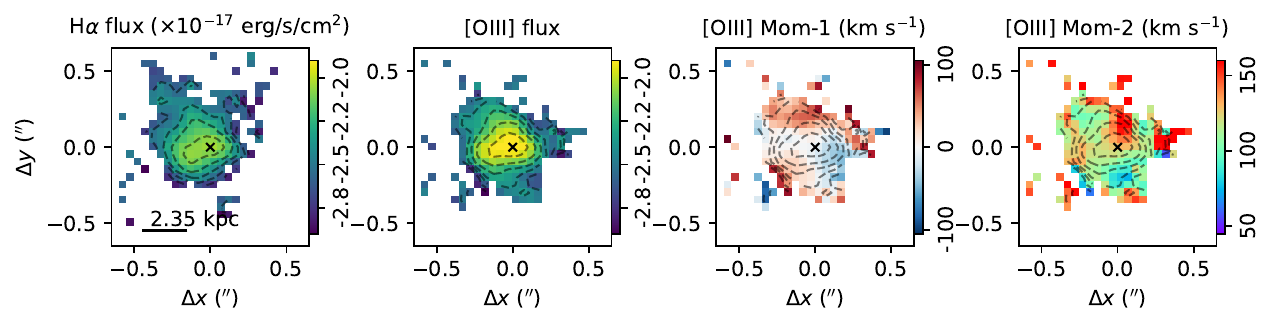}}

    \caption{\ha and \oiii flux distributions, along with the \oiii moment-1 (velocity) and moment-2 (velocity dispersion) maps of LAE2. The flux maps reveal the clumpy morphology of LAE2. The velocity map displays a velocity gradient along the south-east north-west direction in addition to a redshifted component in the northern part of the LAE2, possibly due to an additional component overlapping on the LOS. The contours in the first panel trace \ha emission; those in the remaining panels trace \oiii emission. All images are oriented north up, with east to the left; fluxes are in log-scale.
    }
    \label{fig:resolvedlines}%
    \end{figure*}

Figure~\ref{fig:resolvedlines} displays the spatial distributions of the \ha and \oiii fluxes, as well as the corresponding centroid velocity and velocity-dispersion maps, obtained from spaxel-by-spaxel analysis. Both lines show extended, clumpy morphologies and irregular kinematics, consistent with the presence of multiple substructures already hinted at in the NIRCam F200W imaging.

\subsection{Optical line ratio diagnostics}

\begin{figure*}[!h]
   \centering
   \includegraphics[width=0.45\textwidth, trim=0mm 0mm 0mm 0mm]{{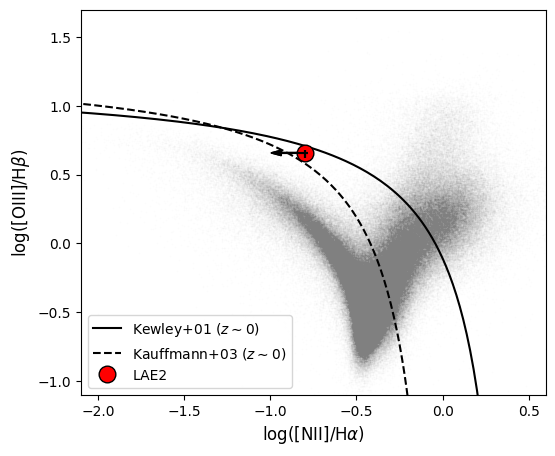}}
   \includegraphics[width=0.45\textwidth, trim=0mm 0mm 0mm 0mm]{{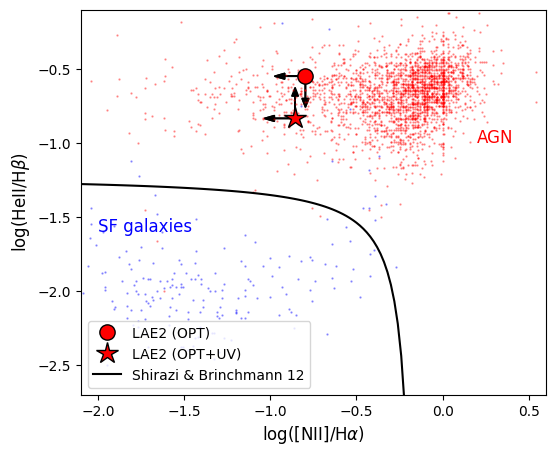}}

    \caption{BPT and \heii diagnostic diagrams. Left: \nii/\ha versus \oiii/\hb for LAE2 (red symbol), compared with local SDSS galaxies (grey points; \citealt{Abazajian2009}). The solid \citep{Kewley2001} and dashed \citep{Kauffmann2003} curves mark the classical boundaries between star-forming galaxies (below the curves) and AGN (above). 
    Right: \nii/\ha versus \heii$\lambda4687$/\hb for LAE2 (large red symbols), and local star-forming galaxies (blue) and AGN (red) from \citet{Shirazi2012}; the solid curve is used to separate star-forming galaxies (below the curve) and AGN (above), according to \citet{Shirazi2012}. The red circle indicates the 3$\sigma$ upper limit on \heii$\lambda4687$/\hb (no \heii$\lambda4687$ detection). The red star shows the corresponding lower limit obtained by assuming a fixed intrinsic ratio \heii$\lambda1640$/\heii$\lambda4687$ = 6.47, given that the UV line is detected but may be affected by extinction. Both symbols share the same \nii/\ha ratio but are slightly offset horizontally for clarity.
    }
    \label{fig:opticaldiagnostics}%
    \end{figure*}

We investigated the dominant ionisation source for the emitting gas in LAE2 using
the classical BPT diagram (\citealt{Baldwin1981}). 
Figure~\ref{fig:opticaldiagnostics} (left) shows the \oiii/\hb versus \nii/\ha flux ratios measured from the nuclear ($r=0.15\arcsec$) spectrum. For comparison, the BPT also displays other optical line ratio measurements from low-$z$ SDSS galaxies (small grey points), and the demarcation lines used to separate galaxies and AGN at $z\sim 0$.

Since the physical conditions of the ISM in galaxies at $z>3$ differ substantially from the local population, both AGN and star-forming galaxies tend to populate the same region of the BPT diagram when associated with low gas metallicities (e.g. \citealt{Feltre2016, Nakajima2022, Harikane2023, Ubler2023, Maiolino2023c}), with low \nii/\ha and high \oiii/\hb, similar to those measured for LAE2. Therefore, additional diagnostics tailored to the early Universe have been proposed, several of which exploit the properties of \heii\!$\lambda$4687 emission \citep[e.g.][]{Shirazi2012, Nakajima2022, Tozzi2023}.
Figure~\ref{fig:opticaldiagnostics} (right) shows LAE2 measurements on the \heii\!$\lambda$4687/\hb versus \nii/\ha diagram. In the rest-optical NIRSpec spectrum both \heii\!$\lambda$4687 and \nii are undetected, yielding an upper limit on both \heii\!$\lambda$4687/\hb and \nii/\ha (large red circle in the figure). Taken alone, this diagram does not tightly constrain the ionisation mechanism because the allowed region overlaps both the AGN and star-forming loci.

However, additional information comes from the \heii\!$\lambda$1640 line detected in the MUSE rest-UV spectrum of LAE2.
Because \heii\!$\lambda$1640 emission is known to be significantly more compact than \lya and other UV lines such as \civ\ or \ciii\ \citep[e.g.][]{Guo2020lya}, it is reasonable to combine the \heii\!$\lambda$1640 flux measured with MUSE (despite its lower spatial resolution) with the \hb flux extracted from our smallest NIRSpec aperture.

Using \textsc{pyneb} (\citealt{Luridiana2015}) to convert the \heii\!$\lambda$1640 flux into the expected optical \heii\!$\lambda$4687 flux taking advantage of their known recombination value \heii\!$\lambda$1640/$\lambda$4687 = 6.47 (assuming an electron temperature of 10$^4$~K, an electron density $n_{\rm e} = 100$~cm$^{-3}$, and no dust extinction), we obtain an estimated ratio \heii\!$\lambda$4687/\hb = 0.16. This value represents a lower limit (red star in Fig.~\ref{fig:opticaldiagnostics}), since any dust attenuation preferentially suppresses the UV line and would increase the intrinsic ratio. Higher electron temperatures would also slightly modify the intrinsic recombination ratio (e.g. \heii\!$\lambda1640/\lambda4687 = 7.35$ at $T_{\rm e}=2\times10^{4}$~K, with negligible dependence on $n_{\rm e}$), but this would translate into a change of $\sim 10$\% in \heii\!$\lambda$4687/\hb,  without affecting our conclusions.


The combined UV–optical constraints therefore place \heii\!$\lambda$4687/\hb in a definite, AGN-only region of the diagnostic plane, incompatible with pure stellar photoionisation. 
This provides strong evidence that LAE2 hosts an AGN, extending the MUSE-based indications reported by \citet[][see their Fig.~12]{Perna2025dual}.

%

\subsection{\lya escape}

Using hydrogen emission lines, it is possible to derive an escape fraction for \lya under certain assumptions. Assuming Case B recombination and a dust-free environment, $n_{\rm e} = 100$~cm$^{-3}$ and T$_{\rm e} = 10000$~K, the intrinsic \lya/\ha ratio is 8.22 (\citealt{Luridiana2015}). Under these conditions, and neglecting the effects of resonant scattering (see below), the \lya escape fraction $f_{\rm esc}$(\lya) can then be calculated as: $f_{\rm esc}$(\lya) $=$ L(\lya)/(8.22~$\times$~L(\ha)).

Using the MUSE \lya flux integrated over an aperture of $r=0.5\arcsec$, matched to the NIRSpec extraction for \ha, we obtained $f_{\rm esc}(\mathrm{Ly}\alpha)=0.15\pm 0.03$ for LAE2. This value is consistent with previous measurements for \lya-emitting galaxies at similar redshift (e.g. \citealt{Hayes2011, Song2014}).
We note, however, that resonant scattering can substantially modify the intrinsic \lya/\ha ratio by increasing the effective path length of \lya photons and thus enhancing dust absorption, as well as by redistributing \lya emission spatially and spectrally through multiple scatterings that can redirect photons out of the line of sight or allow them to escape through low-column-density channels.
Therefore, we have also applied the semi-empirical calibration by
\cite{Sobral2019}, $f_{\rm esc}$(\lya) $= 0.0048\times $EW(\lya)[$\AA^{-1}$]$~\pm 0.05$, to derive an independent estimate of the \lya escape fraction, resulting in $f_{\rm esc}$(\lya) $= 0.27 \pm 0.07$, broadly consistent with the estimate obtained from \lya/\ha.
The EW is computed assuming for LAE2 a continuum level of $1.4\times 10^{-19}$~\ergs~cm$^{-2}~\AA^{-1}$ close to \lya, detected in HST imaging (from SED analysis, Appendix~\ref{sec:Ased};  \citealt{Perna2025dual}), resulting in an EW of 56~$\AA$.

\begin{figure}[!h]
   \centering
   \includegraphics[width=0.45\textwidth, trim=0mm 0mm 0mm 0mm, clip]{{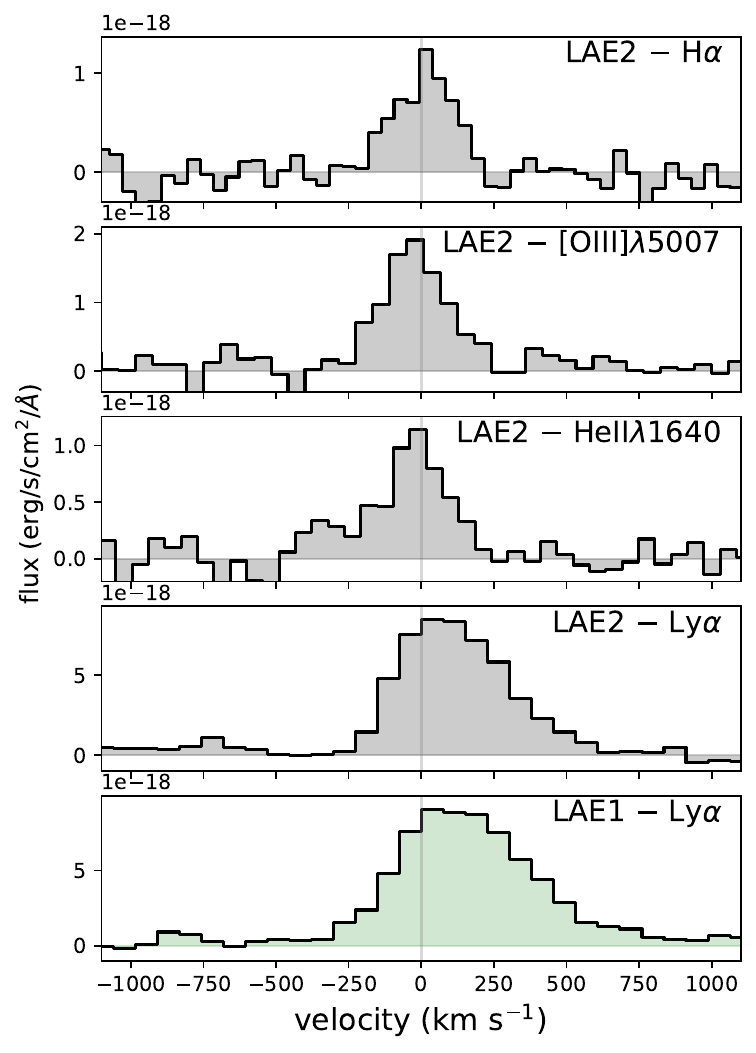}}

    \caption{Emission line profiles of LAE2 (grey) and LAE1 (green) in velocity space. \lya and \heii lines from MUSE spectrum, and \oiii and \ha from NIRSpec IFS, both integrated over an aperture $r = 0.5''$. The vertical line indicates the systemic velocity ($v=0$) of LAE2.
    }
    \label{fig:lineprofilesinvelspace}%
    \end{figure}

For LAE1, we can only place a lower limit on the \lya equivalent width because no continuum is detected. Taking the continuum to be $\lesssim 1/2$ of that of LAE2 (consistent with the HST $\sim 7\sigma$ detection constraints of LAE2 and the non-detection of LAE1) and considering that LAE1 and LAE2 have similar \lya fluxes (Table~\ref{tab:integratedproperties} and Fig.~\ref{fig:lineprofilesinvelspace}), we infer $\mathrm{EW}(\mathrm{Ly}\alpha)\gtrsim 140~\AA$ for LAE1. This approaches the upper-most end of theoretical \lya EWs expected for extremely young stellar populations (i.e. $<10$~Myr, \citealt{CharlotFall1993}). 
Applying the relation of \cite{Sobral2019} then yields a lower limit of $f_{\rm esc}$(\lya)$~\gtrsim 0.67$. An independent constraint can be obtained using the upper limit on the \ha flux from the NIRSpec data, which implies $f_{\rm esc}$(\lya)$~\gtrsim0.64$ from the \lya/\ha ratio. The consistency between these two independent estimates suggests that, under the assumption of in-situ star formation, the \lya escape fraction in LAE1 would be very high.

\section{Discussion}\label{sec:discussion}

The JWST/NIRSpec and VLT/MUSE observations of the \GS-LAE1-LAE2 system provide a rare view of multiple \lya-emitting components embedded within a common extended nebula at $z\sim 3$. The combination of deep NIRSpec and MUSE IFS data reveals markedly different physical conditions in the two LAEs: LAE2 shows high-ionisation UV lines, strong optical nebular emission, bright \lya and faint continuum emission, while LAE1 is detected exclusively in \lya (with a similar luminosity as LAE2) with no continuum counterpart even in the deepest JWST imaging. Understanding the origin of the ionising radiation and the mechanisms powering the \lya emission in these components is essential for reconstructing the physical state of the gas within the system.

In the following subsections, we examine separately the nature of LAE2, whose spectroscopic properties point to AGN-driven ionisation, and LAE1, whose bright but isolated \lya emission raises the question of whether it is powered by scattering, fluorescence, or in-situ star formation.

\subsection{The nature of LAE2}

LAE2 displays an irregular morphology in rest-frame optical emission lines, extending over a few kiloparsecs, with no clear evidence of ordered rotation in its velocity field (Fig.~\ref{fig:resolvedlines}). Rest-frame UV and optical integrated spectra (from MUSE and NIRSpec, respectively) show that \heii\!$\lambda1640$, \oiii\!$\lambda5008$ and \ha  share the same systemic velocity (Fig.~\ref{fig:lineprofilesinvelspace}); this indicates that the UV and optical ionised gas components trace a common kinematic structure. 
The presence of high-ionisation UV transitions and the line ratios inferred from deep MUSE data both point to a hard ionising spectrum \citep{Perna2025dual}. In this work, by combining NIRSpec IFS and MUSE data, we applied both classical and high-redshift optical diagnostics (Fig.~\ref{fig:opticaldiagnostics}); although LAE2 lies outside the regime where BPT diagram is reliable for low-mass and low-metallicity systems, the \heii\!$\lambda4687$ diagnostic robustly places it in the AGN-ionised regime.

The \ha velocity dispersion reaches values of $\approx 100$~\kms (Fig.~\ref{fig:resolvedlines}), which is unusually high for a low-mass, low-SFR system such as LAE2. For comparison, rotation-dominated massive main-sequence galaxies at similar redshifts typically exhibit velocity dispersions of $\sigma \approx 50-60$~\kms, albeit with substantial scatter \citep[e.g.][]{Ubler2019}. Dispersions of order $\sim 100$~\kms are instead observed in systems undergoing extreme star-formation episodes (see e.g. Fig.~7 in \citealt{Perna2022}), well above the star-formation rates inferred for LAE2 from its hydrogen recombination lines (Sect.~\ref{sec:alternativescenarios}). This elevated velocity dispersion is therefore also consistent with the presence of AGN-driven feedback acting on the ionised gas. Alternatively, part of the observed velocity dispersion may arise from the superposition of multiple unresolved clumps (Figs.~\ref{fig:3colours} and \ref{fig:resolvedlines}) and their mutual gravitational interactions.

The \lya spectral profile of LAE2 exhibits a dominant red wing with a small offset from the systemic velocity, $\Delta v_{\rm Ly\alpha} = 105_{-3}^{+5}$~\kms (Fig.~\ref{fig:lineprofilesinvelspace}). 
Radiative-transfer models from \citet{Verhamme2015} provide a useful interpretative framework for this profile. They explored two idealised cases of galaxies leaking Lyman-continuum radiation:
(1) Homogeneous, spherically expanding shells with extremely low H I column densities
(N$_\mathrm{HI} \lesssim 10^{18}$ cm$^{-2}$).
These produce asymmetric redshifted profiles with velocity peaks $\lesssim 150$~\kms from systemic, or double-peaked profiles with peak separations $\lesssim 300$~\kms.
(2) Clumpy shells with incomplete neutral-gas covering, producing profiles peaked very close to systemic.
The \lya profile of LAE2 closely resembles the first class of models, suggesting an interstellar medium dominated by a relatively homogeneous, low-column-density neutral gas component in expansion. This configuration naturally arises when the ISM is exposed to a hard ionising spectrum capable of reducing the effective neutral-gas opacity, also consistent with the presence of an AGN (see also e.g. \citealt{Yang2014}).

In Appendix~\ref{sec:Ased} we present a spectral energy distribution (SED) analysis of LAE2 that extends the work of \citet[][and Circosta et al., in prep.]{Perna2025dual} by explicitly including a potential AGN contribution in the modelling. Allowing for a composite stellar$+$AGN SED, we find that the photometry of LAE2 is well reproduced with an AGN fractional contribution of $f_{\rm AGN}=0.4\pm0.2$. The inclusion of an AGN component leads to a slightly lower inferred stellar mass, $\log(M_\star/\mathrm{M}_\odot)=7.78^{+0.18}_{-0.32}$, compared to purely stellar models. The best-fit solution favours a young stellar population with an age of $\sim$~415~Myr (corresponding to a formation redshift of $z\sim3.5$--4) and a modest ongoing star-formation rate, ${\rm SFR}\approx0.13~\mathrm{M}_\odot\,\mathrm{yr}^{-1}$, in good agreement with the constraints derived from hydrogen recombination lines (Sect.~\ref{sec:alternativescenarios}).

\subsection{The nature of LAE1}

Spatially offset \lya emission with little or no continuum counterpart has been reported in a number of deep narrow-band and IFS studies, and is commonly interpreted as either fluorescence of gas illuminated by an external source, scattering of \lya photons from a nearby galaxy, or cooling radiation from accreting gas (e.g. \citealt{Villar-Martin1996, Bunker2000, Nilsson2006, Geach2009, Cantalupo2012,  Vanzella2017magnifying}).

A number of published systems show qualitative similarities to the LAE1–LAE2 pair.
For instance, \citet{Vanzella2017magnifying} used MUSE observations of a strongly lensed system at $z \simeq 3.3$ to identify a compact \lya emitter located a few kiloparsecs from a bright star-forming galaxy. The \lya line in this offset component is narrow ($\mathrm{FWHM} \lesssim 100~\mathrm{km~s^{-1}}$) and lies at the same systemic redshift as the main galaxy, consistent with fluorescence from gas externally illuminated by escaping radiation from the central source. A faint continuum counterpart is marginally detected in the HST imaging, suggesting a low-mass or gas-dominated clump illuminated by the nearby galaxy.
More recently, \citet{Zarattini2025} reported the discovery of a \lya knot close to a galaxy at $z \simeq 3.5$, characterised by extended (5--8~kpc) \lya emission at a projected distance of a few kiloparsecs. In this case, a faint UV continuum is detected in the knot, indicating that in-situ star formation may contribute to the observed emission, although the authors also discuss the possibility that the structure traces an ionised cone of escaping radiation from the main galaxy. No spectroscopic information is however available for this system.
While morphologically reminiscent of our LAE1–LAE2 system, these examples differ spectroscopically. LAE1 shows a broad and asymmetric \lya profile and remains undetected in the UV, unlike the narrow, symmetric, or continuum-detected knots in the systems mentioned above.
Moreover, a fluorescence origin for LAE1 requires an external source bright enough to power a \lya luminosity of $\sim 10^{42}$ \ergs. Neither LAE2 nor \GS meet this requirement: (a) The X-ray AGN in \GS has an ionising luminosity orders of magnitude too low \citep{Perna2025dual}; (b) \GS is quiescent, with its major star-formation episode ending $\sim 0.5$ Gyr ago; (c) LAE2 has a \lya luminosity comparable to LAE1, thus it cannot power LAE1 through fluorescence at a projected distance of 8 kpc. 
Moreover, if LAE1 were purely fluorescent, \ha emission should be detectable at 2$\sigma$ confidence level, within an aperture of $r = 0.5\arcsec$ assuming a ratio \lya/\ha = 8.22 (or  $>2\sigma$ in smaller apertures). 
Thus, a fluorescence interpretation is strongly disfavoured.

Large \lya blobs (LABs) at high redshift often contain sub-components or `knots' without detectable continuum emission (e.g. \citealt{Herenz2020, Solimano2025}). Some of these knots show broad \lya ($\sigma\approx 200$~\kms) and appear disconnected from known galaxies (e.g. Fig. 11 of \citealt{Jimenez-Andrade2023}).
However, these LAB systems are usually characterised by a \lya nebula extending over $\gtrsim 100$~kpc, global \lya luminosity $>10^{43.5}$~\ergs, and associated with radio galaxies.
Our system \GS-LAE1-LAE2 is less extended ($\sim 30$~kpc), and more balanced in luminosity between components, with L(\lya)~$\approx 10^{42}$~\ergs each, suggesting a different physical configuration than a classical LAB.

LAE1 exhibits a broad and asymmetric \lya profile, no detectable continuum counterpart, and a location embedded within a disturbed \lya halo connecting \GS, LAE2, and LAE1. 
These features are most naturally explained if LAE1 is dominated by resonant scattering of \lya photons produced in LAE2. A roughly similar configuration has been reported in lensed systems at $z = 4$ by \citet{Bunker2000}, in which \lya emission is significantly more extended than the UV continuum and shaped by radiative transfer through neutral gas.

\begin{figure}[!t]
   \centering
   \includegraphics[width=0.48\textwidth, trim=2mm 0mm 2mm 0mm, clip]{{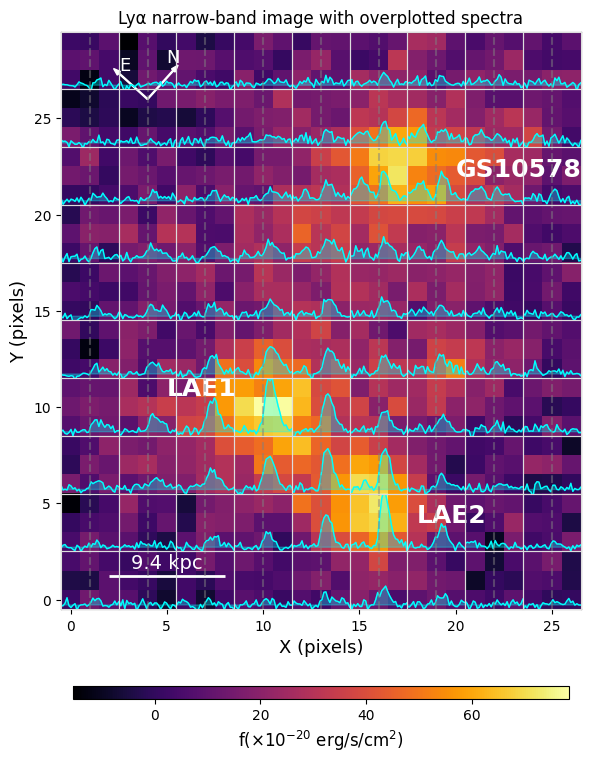}}
   \caption{
   \lya narrow–band image obtained by integrating the MUSE cube over the velocity range [--900, +600]~\kms around the \lya line at the systemic redshift of the \GS galaxy. The background image shows the spatial distribution of the \lya flux, with a compass in the top–left corner indicating the North and East directions and a 1.2\arcsec\ scale bar in the bottom–left. Superimposed are the spectra extracted from $3\times3$–spaxel boxes at the positions marked in the image. All spectra are displayed with the same axis limits, spanning $-1100$ to +1000~\kms, and $(-7$ to $88)\times10^{-20}\ \mathrm{erg~s^{-1}~cm^{-2}~\AA^{-1}}$ along the vertical axis. The vertical dashed line indicates the systemic velocity ($v=0$) of \GS (blueshifted by 200~\kms from LAE2). 
   }
   \label{fig:lyaallspectra}%
\end{figure}

The similarity between the \lya line profiles of LAE1 and LAE2 suggests that both sources are viewed through a common, extended structure of neutral hydrogen, likely an expanding or turbulent screen in the circumgalactic medium that covers both objects. 
The interaction with the massive \GS galaxy may have enhanced the turbulence and bulk motions of this medium, increasing the efficiency of resonant scattering and shaping the emergent \lya profiles.
This is supported by Fig.~\ref{fig:lyaallspectra}, which shows spatially varying \lya profiles over a $\sim40\times40$ kpc$^2$ region: diffuse, perturbed \lya emission is visible between the three emitters, with a double-peaked profile near \GS (peak separation $\sim$600 km s$^{-1}$) and redshifted profile near LAE1 and LAE2.
Within this framework, the markedly different \lya escape fractions and \lya/\ha ratios observed in LAE1 and LAE2 (Table~\ref{tab:integratedproperties}) can be understood if LAE2 hosts an AGN that provides a local source of ionisation, while LAE1 lacks in-situ star formation and primarily reprocesses \lya photons propagating through the shared neutral medium.

\subsection{Alternative scenario for LAE1?}\label{sec:alternativescenarios}

Although the observational evidence favours a scattering-dominated origin for LAE1, it is worth considering whether LAE1 could instead host its own ionising source, hence whether it is a faint galaxy whose stellar continuum remains undetected even in deep JWST data.

To first order, the SFR inferred from recombination lines falls in the regime of low-mass galaxies.
We derive an estimate for the SFR from the \lya line, starting from the better-calibrated \ha star formation indicator (\citealt{Kennicutt2012}), and assuming case-B recombination: SFR(\lya) [\Msunyr] = $5\times 10^{-43}$~L(\lya) [\ergs]. We obtain SFR(LAE1) = 0.4~\Msunyr. 
According to the star-forming main sequence at this redshift \citep{speagle_2014}, such SFRs correspond to stellar masses of order $\lesssim 10^8$~M$_\odot$. 
Thus, if LAE1 hosts in-situ star formation, it would correspond to a very low-mass galaxy. 
In Appendix~\ref{sec:Ased} we used the non-detections in the available HST and JWST imaging to perform a tentative SED analysis and place an upper limit on the stellar mass of LAE1 of M$_\star \lesssim 10^{6.5}$~M$_\odot$. This limit is more than an order of magnitude lower than the stellar mass inferred for LAE2 (M$_\star \approx 10^{8}$~M$_\odot$), and is difficult to reconcile with the SFR implied by the \lya luminosity, and would require very high specific star formation if the \lya emission were powered by in-situ star formation.

Given that the SED constraints alone cannot fully rule out a faint stellar component, we next turn to rest-frame UV emission-line diagnostics to further test the possibility of in-situ star formation in LAE1.
A powerful diagnostic for faint star-forming galaxies at $z\sim~3$ is the \ciii\!$\lambda\lambda1907,1909$ doublet, typically the brightest UV metal line in this population (e.g. \citealt{Shapley2003}).
If LAE1 were powered by its own stars, the empirical correlation between \lya and \ciii equivalent widths \citep{Llerena2022} provides an estimate of the expected \ciii strength.  
Given the lower limit EW(\lya) $> 140$~\AA, the relation predicts EW(\ciii) $\gtrsim 16$~\AA.

Assuming LAE1 has an intrinsic UV slope $\beta=-2.62$ \citep{Boquien2022} and negligible dust, the continuum near \ciii scales as
\begin{equation}
f_{\rm con,\,CIII]} = \left( \frac{\lambda_{\rm CIII]}}{\lambda_{\rm Ly\alpha}} \right)^\beta f_{\rm con,\,Ly\alpha} \approx 0.3\, f_{\rm con,\,Ly\alpha}, 
\end{equation}
where $f_{\rm con, Ly\alpha}$ is the continuum near \lya. Under these assumptions, the expected \ciii line luminosity can be obtained from
\begin{equation}
\frac{\mathrm{EW(Ly}\alpha)}{\mathrm{EW(CIII]})} = \frac{L(\mathrm{Ly}\alpha) / f_{\rm con, Ly\alpha}}{L(\mathrm{CIII]}) / f_{\rm con, CIII]}} = 0.3 
\times \frac{L(\mathrm{Ly}\alpha)}{L(\mathrm{CIII]})},
\end{equation}

To first order, we assumed an EW(\lya) = 140~\AA\ (i.e. similar to the measured lower limit), and take advantage of the known correlation between EW(\lya) and EW(\ciii) for star-forming galaxies at $z\sim 3$ from \citet[][their Eq.~1]{Llerena2022} to obtain an estimate of EW(\ciii) = 16~\AA. By doing so, the previous equation can be used to obtain an estimate of \ciii luminosity, as L(\ciii) = 0.03~$\times$~L(\lya) = 10$^{40.43}$~\ergs.

Given the MUSE noise level near \ciii  of $7 \times 10^{-20}$ erg\,s$^{-1}$\,cm$^{-2}$\,\AA$^{-1}$ and assuming a velocity dispersion of 120~\kms (as for the detected \ciii in LAE2), the $3\sigma$ upper limit is L$^{up}$(\ciii) $= 10^{40.53}$~\ergs.  
Therefore, an extremely low-mass stellar population cannot yet be firmly excluded.

However, we note that if LAE1 were an independent galaxy with its own ionising source, one would expect its \lya profile to reflect its internal ISM conditions (gas kinematics, column density, outflows) and therefore would differ from that of LAE2.  
In contrast, the two profiles share a similar asymmetric shape (Figs.~\ref{fig:lineprofilesinvelspace} and \ref{fig:lyaallspectra}). This behaviour favours a connection between the two \lya-emitters.

\section{Conclusions}\label{sec:conclusions}

We have presented new JWST/NIRSpec IFS observations of the environment of the massive quiescent galaxy \GS at $z=3.064$, targeting two nearby \lya-emitters (LAE1 and LAE2) previously identified in deep VLT/MUSE data. Combined with the MUSE cube and ancillary JWST/NIRCam imaging, our analysis provides a spatially resolved view of ionised gas, kinematics, and AGN activity within $\sim30$~kpc of one of the earliest quenched galaxies known at high redshift. Our main conclusions are as follows:

\begin{itemize}
    \item Our NIRSpec data confirm the presence of an active nucleus in LAE2, adding to the AGN in \GS itself and to the previously identified obscured nucleus in the nearby satellite (both outside the FOV of the NIRSpec data discussed here; \citealt{Perna2025dual}). This system therefore hosts three AGN within $\sim30$~kpc, making it one of the most compact multi-AGN environments known at $z>3$ (see also e.g. \citealt{Perna2025ct,Zamora2025}). This configuration provides additional direct evidence for accelerated black-hole growth in dense environments surrounding massive quenched galaxies.
    \item LAE1 shows bright \lya emission with no detected continuum counterpart in either NIRCam or MIRI imaging. Its broad, asymmetric \lya\ profile, lack of Balmer and metal lines, and spatial correspondence with the \lya\ halo around \GS\ favour an interpretation in which LAE1 traces scattered emission from LAE2, rather than an independent galaxy. 
    A low-mass ($\lesssim 10^{6.5}$~M$_\odot$) stellar population in LAE1 cannot be firmly excluded, but the fact that the \lya profiles in LAE1 and LAE2 are very similar favours a connection between the two. In particular, an expanding or turbulent screen in the circumgalactic medium that covers both objects could shape their emergent \lya emission.
\end{itemize}

The environment of \GS is a site of ongoing gas accretion and black-hole growth.
Despite \GS being fully quenched for $\sim0.5$~Gyr, its surroundings host gas-rich satellites, extended \lya nebula, and multiple AGN. This suggests that massive quenched galaxies at early times can reside in dynamically complex and gas-rich environments where SMBH fuelling continues independently of the massive galaxy's star-formation history.

Overall, the \GS-LAE1-LAE2 system highlights the power of combining JWST IFS data with deep MUSE spectroscopy to dissect the interplay of AGN activity, gas accretion, and galaxy evolution at $z>3$. Future wide-area surveys and deep JWST observations will be essential to determine how common such compact multi-AGN environments are, and to establish their role in shaping the early growth of massive galaxies and their central SMBHs.

\begin{acknowledgements}


We thank Luis Colina, Manuel Solimano, Lorenzo Ulivi, and Eros Vanzella for valuable discussions.

MP, SA, BRP, and PPG acknowledge support from the research projects PID2021-127718NB-I00, PID2024-159902NA-I00, PID2024-158856NA-I00, and RYC2023-044853-I of the Spanish Ministry of Science and Innovation/State Agency of Research (MCIN/AEI/10.13039/501100011033) and FSE+.
IL acknowledges support from PRIN-MUR project “PROMETEUS”  financed by the European Union -  Next Generation EU, Mission 4 Component 1 CUP B53D23004750006.
RM acknowledges support by the Science and Technology Facilities Council (STFC), by the ERC Advanced Grant 695671 ``QUENCH'', and by the UKRI Frontier Research grant RISEandFALL; RM is further supported by a research professorship from the Royal Society.
AJB acknowledges funding from the ``First Galaxies'' Advanced Grant from the European Research Council (ERC) under the European Union’s Horizon 2020 research and innovation programme (Grant agreement No. 789056).
%
%
SC and GV acknowledge support by European Union’s HE ERC Starting Grant No. 101040227 - WINGS.
MP, GC and EB acknowledge the support of the INAF Large Grant 2022 "The metal circle: a new sharp view of the baryon cycle up to Cosmic Dawn with the latest generation IFU facilities". 
GC and EB also acknowledge the INAF GO grant ``A JWST/MIRI MIRACLE: Mid-IR Activity of Circumnuclear Line Emission''. EB acknowledges funding through the INAF ``Ricerca Fondamentale 2024'' programme (mini-grant 1.05.24.07.01).

H\"U acknowledges funding by the European Union (ERC APEX, 101164796). Views and opinions expressed are however those of the authors only and do not necessarily reflect those of the European Union or the European Research Council Executive Agency. Neither the European Union nor the granting authority can be held responsible for them.
\end{acknowledgements}

\bibliographystyle{aa}
\bibliography{aanda_1stsub.bib}

\begin{appendix}
\onecolumn 
\section{Spectral energy distribution of LAE2 and LAE1}\label{sec:Ased}

\begin{figure*}[]
   \centering
   \includegraphics[width=1\textwidth]{{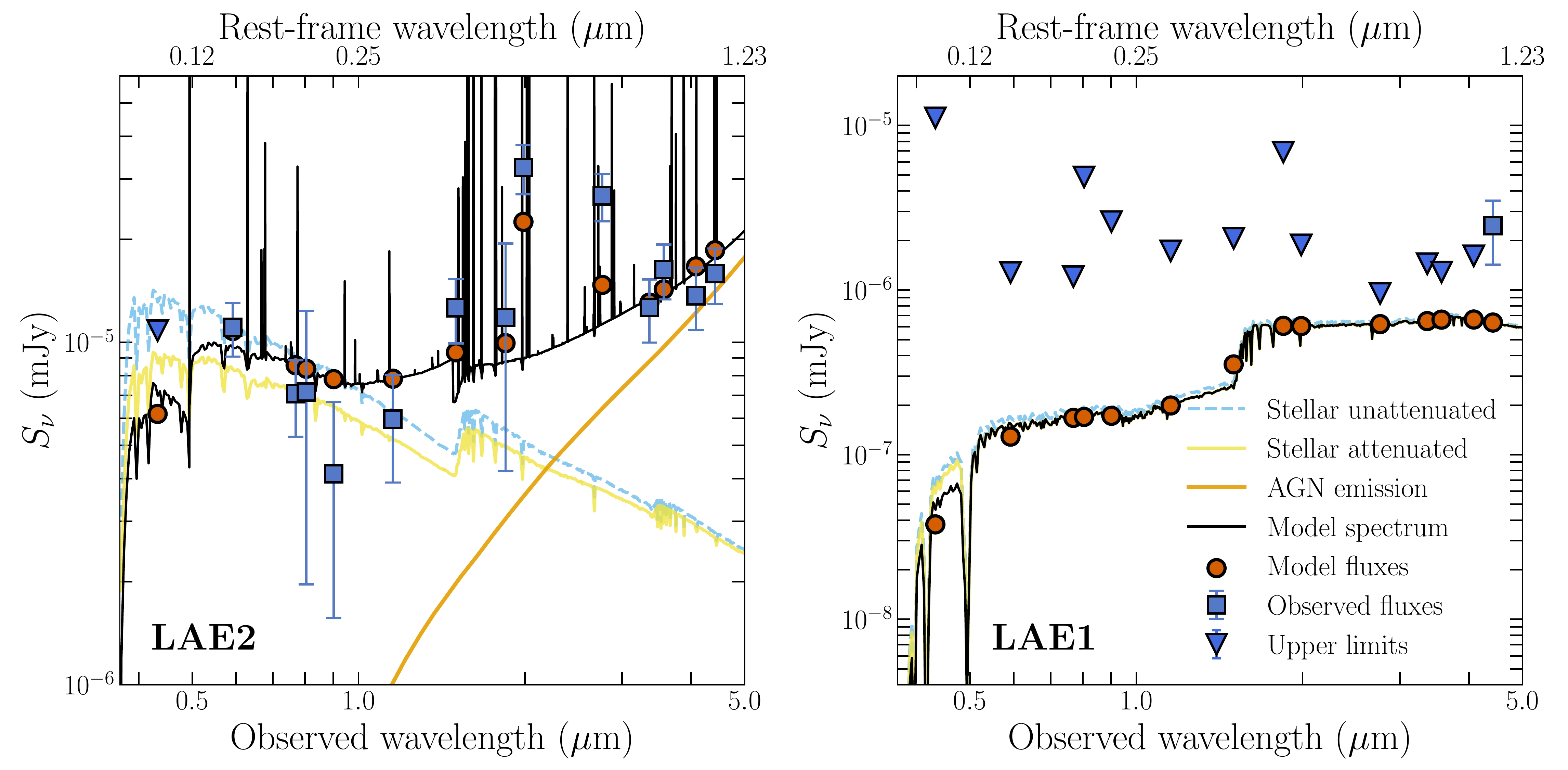}}
   \caption{Best-fit spectral energy distributions for LAE2 (left) and LAE1 (right), computed using the configuration described in \ref{sec:Ased}. The observed photometry (blue squares for detections, blue triangles for $3\sigma$ upper limits) spans HST/ACS through JWST/NIRCam. For LAE1, the reddest point is marked with a square, because used as an anchor point for the SED fit, but represents an upper limit. Model fluxes are shown as red circles. The solid black line shows the total model spectrum, while colored lines indicate individual components. The top axis shows rest-frame wavelength at $z = 3.07$.}
   \label{fig:SEDs}%
\end{figure*}

To characterise the stellar populations of LAE2 and 
explore potential constraints on LAE1, we attempted SED fitting using the available broadband photometry from archival HST and JWST imaging. We used the available archival HST (ACS/WFC F435W, F606W, F775W, F814W) and JWST/NIRCam imaging (F090W, F115W, F150W, F182M, F200W, F277W, F335M, F356W, F410M, F444W) from the JADES survey \citep{Eisenstein2023, Rieke2023, Eisenstein2025}. We PSF-matched all maps to the reddest filter (NIRCam's F444W band, with a measured PSF FWHM of 0.16$\arcsec$) to enable consistent aperture photometry across wavelengths. We masked a nearby star located $\sim$2$\arcsec$ from the system prior to photometric extraction to reduce contamination.We did not include JWST/MIRI data in the SED analysis, as both LAE2 and LAE1 remain undetected in these bands, and the relatively shallow MIRI exposures (\citealt{Alberts2024SMILES}) do not provide meaningful additional constraints on the fits.

For LAE2, we extracted aperture photometry using an elliptical aperture optimized to capture the source's morphology while maximizing S/N. LAE2 is detected in 13 filters (average S/N $\sim$ 7.5), including 7 strong detections (S/N $\sim$ 11) and 6 tentative detections (S/N $\sim$ 3). For LAE1, located 0.9$\arcsec$ ($\sim$7 kpc) east from LAE2, we applied a similar elliptical aperture centred on the position of the Ly$\alpha$ emission identified in MUSE for LAE1. Despite the depth of the JWST imaging, LAE1 shows no continuum detections in any filter, 
yielding only upper limits across all bands.

We fitted the observed photometry using the Code Investigating GALaxy Emission (\texttt{CIGALE}; \citealt{Boquien2019}), a Bayesian SED fitting code that models panchromatic galaxy emission. We first fitted LAE2, given its multiple continuum detections, and then attempted to constrain LAE1 despite the lack of detections.

We adopted model components and priors tailored to the AGN-hosting nature of LAE2. For the star formation history (SFH), we used delayed exponentially declining models with $\text{SFR}(t) \propto t \exp(-t/\tau)$, with e-folding times $\tau$ kept as free parameter. Stellar populations were modelled using \cite{bruzual_2003} templates with \cite{Chabrier2003} IMF and sub-solar metallicities. Nebular emission was computed using Cloudy models with ionization parameter log$U$ ranging from -3 to -1 (spanning star-forming to AGN-dominated ionization regimes). We fixed the nebular line widths to 112 km s$^{-1}$ based on our spectroscopic measurements (Table~\ref{tab:integratedproperties}). The Lyman continuum escape fraction was allowed to vary from 0 to 0.3, consistent with the constraints derived from the \lya/\ha ratio (\ref{sec:results}).

Dust attenuation was modelled using a modified \cite{Calzetti2000} law with colour excess E(B-V) ranging from 0 to 0.03~mag, appropriate for low-attenuation systems like LAE2. We allowed the slope of the attenuation curve to vary between -2 and 2. Given the confirmed AGN nature of LAE2 (Sect.~\ref{sec:results}), we included \cite{Fritz2006} clumpy torus models in our SED fitting analysis. We sampled optical depths at 9.7 microns ranging from 0.3 to 10, torus opening angles between 60 and 140 degrees, and three viewing angles corresponding to face-on, intermediate, and edge-on configurations (Type 1, intermediate, and Type 2 AGN). We adopted the \cite{Schartmann2005} disk spectrum, and dust temperatures between 100 and 300 K.

For LAE1, given the absence of continuum detections, we adopted a different strategy to enable SED fitting. We treated the reddest band (F444W, S/N $\sim$ 2) as a detection to provide a minimal anchor point for \texttt{CIGALE}, while all other filters were kept as upper limits. We applied the same model components and priors as for LAE2, but excluded the AGN scenario since with only upper limits, we can not constrain complex models. Although attempting to forge an SED for LAE1 is speculative, it allows us to derive upper limits on the assumed stellar mass and SFR. Given the lack of detections, the resulting SED fit represents stringent constraints rather than physical measurements.\\

\begin{table}
\centering
\caption{Physical properties from SED fitting for LAE2 and LAE1.}
\begin{tabular}{lcc}
\hline\hline
Property & LAE2 & LAE1 \small(upper limits) \\
\hline
$\log(M_\star/M_\odot)$ & 7.78$^{+0.18}_{-0.32}$ & 6.19$^{+0.18}_{-0.32}$ \\
SFR [$M_\odot$ yr$^{-1}$] & 0.13 $\pm$ 0.07 & 0.02 $\pm$ 0.03 \\
$f_{\rm AGN}$ & 0.40 $\pm$ 0.17 & --- \\
Age [Myr] & 416 $\pm$ 162 & 158 $\pm$ 55 \\
$A_V$ [mag] & 0.07 $\pm$ 0.05 & 0.02 $\pm$ 0.01 \\
$Z$ [$Z_\odot$] & 0.16 $\pm$ 0.13 & 0.14 $\pm$ 0.09 \\
$\log U$ & -1.39 $\pm$ 0.49 & -2.50 $\pm$ 0.41 \\
\hline
\end{tabular}
\tablefoot{Errors represent 68\% confidence intervals (16th-84th percentiles). LAE2 was fitted with stellar+AGN model; LAE1 with stellar-only model. LAE1 results represent upper limits given the lack of continuum detections.}
\label{tab:sed_results}
\end{table}

The computed SED for LAE2 and LAE1 are shown in Figure \ref{fig:SEDs}, and the resulting physical properties corresponding to these best fit solutions are shown in Table \ref{tab:sed_results}. The SED fitting results confirm that LAE2 is a low-mass galaxy (log(M$_\star$/M$_\odot$) = 7.78$^{+0.18}_{-0.32}$, with a young stellar population (age $\approx$ 415 Myr, corresponding to formation at $z \sim 3.5$--4) and modest ongoing star formation (SFR $\approx$ 0.13 M$_\odot$ yr$^{-1}$, consistent with estimates based on Balmer lines in Sect.~\ref{sec:alternativescenarios}). The derived AGN fraction of $f_{AGN} = 0.4 \pm 0.2$ confirms that AGN emission contributes significantly to the total infrared luminosity. For LAE1, the SED fit yields only weak constraints given the absence of continuum detections. The inferred stellar mass upper limit (log(M$_\star$/M$_\odot$) $\lesssim$ 6.5) and very low SFR ($\lesssim$ 0.02 M$_\odot$ yr$^{-1}$) are consistent with LAE1 being either an extremely low-mass, metal-poor dwarf galaxy or dominated by scattered Ly$\alpha$ emission with no stellar contribution, as suggested by the spectroscopic analysis.

\end{appendix}

\end{document}